\documentclass[onecolumn,showpacs,showkeys,amsmath,amssymb,superscriptaddress,floatfix,nofootinbib]{revtex4}

\usepackage{graphicx}
\usepackage{bm} 
\usepackage{subfigure}
\usepackage[colorlinks=true,linktocpage=true,linkcolor=blue,citecolor=blue]{hyperref}
\usepackage[usenames,dvipsnames]{color}

\makeatletter
\newcommand\erfc{\mathop{\operator@font erfc}\nolimits}
\def\slashchar#1{\setbox0=\hbox{$#1$}
   \dimen0=\wd0 \setbox1=\hbox{/} \dimen1=\wd1
   \ifdim\dimen0>\dimen1 \rlap{\hbox to \dimen0{\hfil/\hfil}} #1
   \else  \rlap{\hbox to \dimen1{\hfil$#1$\hfil}} / \fi}

\newcommand{\taueq}{\tau_{\rm eq.}}

\newcommand{\tildeN}{{\tilde N}}

\newcommand{\ptrans}{p_\perp}
\newcommand{\plong}{p_\|}
\newcommand{\sigmatrans}{\sigma_\perp}
\newcommand{\sigmalong}{\sigma_\|}
\newcommand{\Thetatrans}{\Theta_\perp}
\newcommand{\Thetalong}{\Theta_\|}
\newcommand{\alphatrans}{\alpha_\perp}
\newcommand{\alphalong}{\alpha_\|}

\newcommand{\Thetaeq}{\Theta_{\rm eq.}}
\newcommand{\Thetatreq}{\Thetatr^{\rm eq.}}

\newcommand{\mh}{{\hat m}}
\newcommand{\mheq}{\mh_{\rm eq.}}
\newcommand{\pres}{{\cal P}}
\newcommand{\preseq}{\pres_{\rm eq.}}
\newcommand{\ped}{{\cal E}}
\newcommand{\pedeq}{\ped_{\rm eq.}}

\newcommand{\Thetatr}{\Theta_{\sf tr}}

\newcommand{\mhat}{{\hat m}}

\begin{document}
 
\title{Testing different formulations of leading-order anisotropic hydrodynamics}

\author{Leonardo Tinti} 

\affiliation{Institute of Physics, Jan Kochanowski University, PL-25406~Kielce, Poland} 

\author{Radoslaw Ryblewski} 

\affiliation{The H. Niewodnicza\'nski Institute of Nuclear Physics, Polish Academy of Sciences, PL-31342~Krak\'ow, Poland} 

\author{Wojciech Florkowski} 

\affiliation{The H. Niewodnicza\'nski Institute of Nuclear Physics, Polish Academy of Sciences, PL-31342~Krak\'ow, Poland} 

\author{Michael Strickland} 

\affiliation{Department of Physics, Kent State University, Kent, Ohio 44242, USA}

\begin{abstract}
A recently obtained set of the equations for leading-order (3+1)D anisotropic hydrodynamics is tested against exact solutions of the Boltzmann equation with the collisional kernel treated in the relaxation time approximation. In order to perform the detailed comparisons, the new anisotropic hydrodynamics equations are reduced to the boost-invariant and transversally homogeneous case.  The agreement with the exact solutions found using the new anisotropic hydrodynamics equations is similar to that found using previous, less general, formulations of anisotropic hydrodynamics.  In addition, we find that, when compared to a state-of-the-art  second-order viscous hydrodynamics framework, leading-order anisotropic hydrodynamics better reproduces the exact solution for the pressure anisotropy and gives comparable results for the bulk pressure evolution.  Finally, we compare the transport coefficients obtained using linearized anisotropic hydrodynamics with results obtained using second-order viscous hydrodynamics.
\end{abstract}



\maketitle 

\section{Introduction}
\label{sect:intro}

The successful explanation of the space-time evolution of matter produced in relativistic heavy-ion collisions at RHIC (Relativistic Heavy-Ion Collider) and the LHC (Large Hadron Collider) using relativistic viscous hydrodynamics initiated new developments in the hydrodynamics
\cite{Israel:1976tn,Israel:1979wp,
Muronga:2001zk,Muronga:2003ta,
Baier:2006um,Baier:2007ix,
Romatschke:2007mq,Dusling:2007gi,Luzum:2008cw,
Song:2008hj,El:2009vj,PeraltaRamos:2010je,
Denicol:2010tr,Denicol:2010xn,
Schenke:2010rr,Schenke:2011tv,
Bozek:2009dw,Bozek:2011wa,
Niemi:2011ix,Niemi:2012ry,
Bozek:2012qs,Denicol:2012cn,Jaiswal:2013npa,Denicol:2014loa}. 
An example of such an activity is the burgeoning study of {\it anisotropic hydrodynamics}. In its original formulation,  anisotropic hydrodynamics was restricted to boost-invariant and transversally homogeneous systems \cite{Florkowski:2010cf,Martinez:2010sc}, denoted below as the (0+1)D case. Shortly afterwards, this approach was generalised to the one-dimensional, non boost-invariant situation \cite{Ryblewski:2010bs,Martinez:2010sd}. The (2+1)D and (3+1)D  cases (corresponding to a two-dimensional transverse expansion with longitudinal boost-invariance and to an arbitrary three-dimensional expansion without any symmetry constraints, respectively) were studied in a subsequent series of papers \cite{Ryblewski:2011aq,Martinez:2012tu,Ryblewski:2012rr,Ryblewski:2013jsa}. In parallel, the concept of anisotropic hydrodynamics was extended to describe mixtures of anisotropic quark and gluon fluids \cite{Florkowski:2012as,Florkowski:2013uqa,Florkowski:2014txa}.

In Refs.~\cite{Florkowski:2010cf,Martinez:2010sc,Ryblewski:2010bs,Martinez:2010sd,Ryblewski:2011aq,Martinez:2012tu,Ryblewski:2012rr,Ryblewski:2013jsa,Florkowski:2012as,Florkowski:2013uqa,Florkowski:2014txa} only the leading order of the hydrodynamic expansion has been included  and described by the spheroidal Romatschke-Strickland form \cite{Romatschke:2003ms}. This form allows for the difference between the longitudinal and transverse pressures only (with the cylindrical symmetry in the transverse plane) and hence cannot describe more complex situations where, for example,  the three components of pressure differ from each other.  A successful method to solve this problem by including further terms in the hydrodynamic expansion was first  presented in Ref.~\cite{Bazow:2013ifa} for massless (conformal) systems and was recently extended to massive (non-conformal) systems in Ref.~\cite{Bazow:2015cha}.

An alternative starting point for performing the anisotropic hydrodynamics expansion is to include the momentum anisotropy in a more complete way already at leading order by generalizing the Romatschke-Strickland form to allow the energy-momentum tensor to possess different pressures along all three spatial directions in the local rest frame (ellipsoidal form). Such a non-perturbative treatment for the (1+1)D case (boost-invariant expansion with cylindrical spatial symmetry) was proposed  for conformal systems in Ref.~\cite{Tinti:2013vba} and soon generalized to non-conformal systems in Ref.~\cite{Nopoush:2014pfa}.  Recently, a more general set of equations for leading-order anisotropic hydrodynamics has been presented by Tinti in Ref.~\cite{Tinti:2014cpa}. Differently from previous leading-order approaches, the new formulation does not use any simplifying symmetry assumptions such as the longitudinal boost invariance or cylindrical symmetry. 

Using this new approach it was shown that, in the close-to-equilibrium limit, the general anisotropic hydrodynamics equations have the same structure as the corresponding second-order viscous hydrodynamics equations \cite{Tinti:2014cpa}. However, in the process one finds that anisotropic hydrodynamics allows for several different ways to obtain the equations necessary to describe the evolution of the bulk pressure of the system.  In order to determine which option is the best for possible phenomenological applications, in this paper we test the corresponding solutions with the help of exact solutions of the Boltzmann equation treated in the relaxation time approximation.  To perform the numerical comparisons, the new equations are reduced to the boost-invariant and transversally homogeneous case, where the solutions of the Boltzmann equation are available.  We note that comparisons of the predictions of various hydrodynamic frameworks with the solutions of the underlying kinetic theory has now become a quite useful method for testing the accuracy of different hydrodynamic formulations \cite{Florkowski:2013lza,Florkowski:2013lya,Denicol:2014mca,Jaiswal:2014isa,Denicol:2014xca,Denicol:2014tha,Nopoush:2014qba}.

In this paper, we show that the agreement of the new formulation of anisotropic hydrodynamics with the exact solutions is similar to that obtained using earlier, less general, formulations of anisotropic hydrodynamics.  We find that leading-order anisotropic hydrodynamics better reproduce the exact solution for the pressure anisotropy and gives comparable results for the bulk pressure evolution.  With regard to which method to use to describe the evolution of the bulk viscous pressure, we find that it is better to use the zeroth moment of the kinetic equation, while for the ratio of the longitudinal to transverse pressure it is best to use the trace of the second moment instead of the zeroth moment.

The paper is organized as follows: In Sec.~\ref{sect:lorder}, we present the equations of leading-order (3+1)D anisotropic hydrodynamics derived originally in Ref.~\cite{Tinti:2014cpa} --- we first define the energy-momentum tensor in the Landau frame, then the phase space distribution is introduced and discussed, and finally, the forms of the hydrodynamic equations obtained from the first three moments of the Boltzmann equation are presented. In Sec.~\ref{sect:0+1}, the symmetry conditions characterizing boost-invariant and transversally-homogeneous systems are implemented and special forms of various physical quantities in the (0+1)D case are presented. In Sec.~\ref{sect:reducedeq}, we present the leading-order anisotropic hydrodynamics equations reduced to the (0+1)D case. In Sec.~\ref{sect:num}, we present our numerical results, first for the fully non-perturbative case (Sec.~\ref{sect:num1}), and then for the linearized case (Sec.~\ref{sect:num2}). We summarize and conclude in Sec.~\ref{sect:con}.

\section{Leading order (3+1)D anisotropic hydrodynamics}
\label{sect:lorder}

\subsection{Energy-momentum tensor}

In this Section we recall the general structure of the equations of (3+1)D leading-order anisotropic hydrodynamics as introduced recently by Tinti \cite{Tinti:2014cpa}. In this case, one adopts the Landau frame and, consequently, uses the following form of the energy-momentum tensor
\begin{equation}
 T^{\mu\nu} = T_{\rm eq.}^{\mu\nu} + \pi^{\mu\nu} - \Pi \Delta^{\mu\nu} \quad 
= \quad \ped \, U^\mu U^\nu - \left(\vphantom{\frac{}{}} \preseq +\Pi \right) \Delta^{\mu\nu} + \pi^{\mu\nu} .
\label{Tmunu_decomposition}
\end{equation}
Here the tensor $ T_{\rm eq.}^{\mu\nu}$ describes the equilibrium part of $ T^{\mu\nu}$, which has the perfect-fluid form
\begin{equation}
 T^{\mu\nu}_{\rm eq.} = \pedeq U^\mu U^\nu - \preseq \Delta^{\mu\nu} ,
\end{equation}
expressed by the equilibrium energy density, 
\begin{equation}
\pedeq = 4\pi \tildeN \, T^4 \mheq^2 \left[  3K_2(\mheq)+ \mheq K_1(\mheq)\right],
\label{Eeq}
\end{equation}
the equilibrium pressure $\preseq$, 
\begin{eqnarray}
\preseq = 4\pi\tildeN \, T^4\mheq^2 K_2(\mheq), 
\label{Peq}
\end{eqnarray}
and the flow vector $U^\mu$. In Eqs.~(\ref{Eeq}) and (\ref{Peq}) $\tilde{N} = N_{\rm dof}/(2\pi)^3$, with $N_{\rm dof}$ being the number of internal degrees of freedom, $\mheq = m/T$, where $m$ is the particle mass and $T$ is the temperature.  We note that Eqs.~(\ref{Eeq}) and (\ref{Peq}) describe a classical massive gas using a Boltzmann distribution. Such a system has been recently studied also in Refs.~\cite{Florkowski:2014sfa,Florkowski:2014bba,Nopoush:2014pfa}. Generalizations to the case of quantum (Bose-Einstein or Fermi-Dirac) statistics case can be found, e.g.,  in Refs.~\cite{Florkowski:2014sda,Florkowski:2015lra}. 

The projection tensor $\Delta^{\mu\nu}$ is defined as
\begin{equation}
 \Delta^{\mu\nu} = g^{\mu\nu} - U^\mu U^\nu.
\label{Delta}
\end{equation}
The quantity $\Pi$ is the bulk viscous pressure, while $\pi^{\mu\nu}$ is the shear tensor. For conformal systems (in particular for a gas of massless particles) the bulk viscous pressure vanishes. On the other hand, for non-conformal systems, the bulk viscous pressure is finite and there are substantial couplings between the bulk pressure and shear tensor~\cite{Denicol:2014mca,Jaiswal:2014isa}.

\subsection{The underlying phase-space distribution function and Landau matching}

The energy-momentum tensor $T^{\mu \nu}$ is the second moment of the phase-space distribution function $f(x,p)$, which appears if we analyze the first moment of the Boltzmann equation,
\begin{eqnarray}
T^{\mu \nu} = \int dP \,p^\mu p^\nu\, f(x,p).
\label{Tmunu}
\end{eqnarray}
Here $dP$ is the Lorentz-invariant momentum integration measure $dP = d^3p/E_{\bf p}$, where $E_{\bf p} = \sqrt{{\bf p}^2 + m^2}$ is the particle energy. In the present formulation of anisotropic hydrodynamics we assume that the distribution function is of the form~\cite{Martinez:2012tu,Nopoush:2014pfa}
\begin{equation}
 f = \tildeN \exp\!\left[ -\frac{1}{\lambda}\sqrt{ \frac{}{} p_\mu \Xi^{\mu\nu} p_\nu } \right].
\label{a-Hydro_distribution_0}
\end{equation}
Here, the tensor $\Xi^{\mu\nu}$ reads~\cite{Nopoush:2014pfa}
\begin{equation}
 \Xi^{\mu\nu} = U^\mu U^\nu + \xi^{\mu\nu} -\phi \Delta^{\mu\nu},
\label{decomposition}
\end{equation}
where $\xi^{\mu\nu}$ is the spatial, traceless part of $\Xi^{\mu\nu}$, and  $\phi$ is a degree of freedom controlling the trace of $\Xi^{\mu\nu}$.  The expression (\ref{a-Hydro_distribution_0}) is a generalized Romatschke-Strickland form \cite{Romatschke:2003ms}.

The equilibrium distribution function is recovered in the limit of vanishing anisotropy tensor $\xi^{\mu\nu}$ and vanishing bulk parameter $\phi$. In this case, one has
\begin{equation}
f  =  f_{\rm eq.} = \tildeN \exp\!\left[ -\frac{p_\mu U^\mu}{T} \right].
\label{feq}
\end{equation}
For any non-equilibrium phase-space distribution function $f$, one can find the corresponding thermal distribution $ f_{\rm eq.}$ by requiring that the two distributions yield the same energy density, namely
\begin{eqnarray}
\ped = \int dP (p \cdot U)^2 f(x,p) = \int dP (p \cdot U)^2  f_{\rm eq.}(x,p) = \pedeq.
\label{Landau_matching}
\end{eqnarray}
Equation (\ref{Landau_matching}) is known as the Landau matching. It allows us to determine the effective temperature of the system locally for any non-equilibrium situation. We note that in local equilibrium,  the momentum scale $\lambda$ is equivalent to the temperature $T$. 

\subsection{First-moment equations}

Similarly to other hydrodynamic prescriptions, the first four equations of (3+1)D anisotropic hydrodynamics  follow from energy-momentum conservation
\begin{equation}
 \partial_\mu T^{\mu\nu} = 0.
\label{energy_momentum_conservation}
\end{equation}
We call these equations the first-moment equations as they are obtained from the first moment of the Boltzmann equation. In the relaxation time approximation, Eq.~(\ref{energy_momentum_conservation}) follows from the Landau matching condition (\ref{Landau_matching}).

Instead of using (\ref{energy_momentum_conservation}) directly, we consider the projection of (\ref{energy_momentum_conservation}) on the four-velocity $U$, 
\begin{equation}
 D\pedeq = -\left( \pedeq +\preseq +\Pi \vphantom{\frac{}{}} \right)\theta + \pi^{\mu\nu} \sigma_{\mu\nu},
\label{energy_conservation}
\end{equation}
and the contraction of (\ref{energy_momentum_conservation})  with the projector $\Delta$, which gives
\begin{equation}
 \left( \pedeq + \preseq + \Pi \vphantom{\frac{}{}}\right) DU^\mu = \nabla^\mu \preseq +\nabla^\mu \Pi +\Delta^\mu_\rho \partial_\sigma \pi^{\rho\sigma}.
\label{momentum_conservation}
\end{equation}
In Eq.~(\ref{energy_conservation})~the symbol  $D = U^\mu \partial_\mu$ is the convective derivative (i.e., the time derivative along the fluid flux lines), $\theta$ is the expansion scalar  $\theta=\partial_\mu U^\mu$, and  $\sigma^{\mu\nu} = \partial^{\langle\mu}U^{\nu\rangle}$ is the velocity stress tensor. The angular brackets denote the following tensor structure
\begin{equation}
 {\sf A}^{\langle\mu\nu\rangle} = \Delta^{\mu\nu}_{\alpha\beta} {\sf A}^{\alpha\beta} = \frac{1}{2}\left (\Delta^\mu_\alpha\Delta^\nu_\beta+\Delta^\nu_\alpha\Delta^\mu_\beta-\frac{2}{3}\Delta^{\mu\nu}\Delta_{\alpha\beta}\right){\sf A}^{\alpha\beta}.
\label{Deltamunualphabeta}
\end{equation}
The operator $\nabla^\mu$ appearing on the right-hand side of Eq.~(\ref{momentum_conservation})  is the spatial gradient $\nabla^\mu = \Delta^{\mu\nu}\partial_\nu$. The latter can be used in the standard decomposition of the four velocity gradient
\begin{equation}
 \nabla_\mu U_\nu = \sigma_{\mu\nu} + \frac{1}{3}\theta\Delta_{\mu\nu} +\omega_{\mu\nu},
\end{equation}
where $\omega_{\mu\nu} = \frac{1}{2} ( \nabla_{\mu}U_{\nu} - \nabla_{\nu}U_{\mu}) \equiv \nabla_{[\mu}U_{\nu]}$ is the vorticity.

\subsection{Second and zeroth moment equations}

The analysis of the second moment of the Boltzmann equation performed in \cite{Tinti:2014cpa}  shows that it is convenient to use the tensor 
\begin{equation}
 \Theta^{\mu\nu} = \Delta^\mu_\alpha\Delta^\nu_\beta \int dP \, (p\cdot U) p^\alpha p^\beta f.
\end{equation}
By taking traceless part of $ \Theta^{\mu\nu}$, we obtain the tensor $\Theta^{\langle\mu\nu\rangle} $, which satisfies the following differential equation~\cite{Tinti:2014cpa}
\begin{equation}
 D\Theta^{\langle\mu\nu\rangle} + \frac{1}{\taueq} \Theta^{\langle\mu\nu\rangle}= - 2\Theta^{\langle\mu}_\lambda  \sigma^{\nu\rangle\lambda} - \frac{5}{3}\theta \, \Theta^{\langle\mu\nu\rangle}  + 2\Theta^{\langle\mu}_\lambda \omega^{\nu\rangle\lambda},
\label{Shear_1}
\end{equation}
where $\taueq$ is the relaxation time appearing in the Boltzmann kinetic equation written in the relaxation time approximation \cite{Bhatnagar:1954zz}. We note that, in the close-to-equilibrium case, the tensor $\Theta^{\langle\mu\nu\rangle}$ becomes proportional to the shear tensor $\pi^{\mu\nu}$.  Besides the tensor equation (\ref{Shear_1}), which describes effects related to the shear viscosity, in this paper we will consider three prescriptions for obtaining the additional scalar equation necessary to describe the evolution of the bulk parameter $\phi$.  

\subsubsection{Trace equation}

The first possibility we consider results from taking the trace of the second moment of the Boltzmann equation which gives~\cite{Tinti:2014cpa}
\begin{equation}\label{bulk_2}
D\Thetatr + \frac{5}{3} \, \theta \, \Thetatr - 2 \Theta_{\langle\mu\nu\rangle}\sigma^{\mu\nu} = \frac{1}{\taueq}\left[ \vphantom{\frac{}{}} \Thetatreq - \Thetatr  \right],
\end{equation}
where
\begin{equation}
 \Thetatr= -\Delta_{\alpha\beta}\Theta^{\alpha\beta}.
\end{equation}
%

\subsubsection{U projection}

The second possibility we consider is obtained by projecting the second moment of the Boltzmann equation two times on the flow vector $U$. In this way one obtains~\cite{Tinti:2014cpa}
\begin{eqnarray}\label{UmuUnu_1}
  m^2\left(\vphantom{\frac{}{}} Dn + n \theta \right) +\left[ D\Thetatr + \frac{5}{3} \, \theta \, \Thetatr - 2 \Theta_{\langle\mu\nu\rangle}\sigma^{\mu\nu}\right] = \frac{1}{\taueq}\left\{ m^2\left[\vphantom{\frac{}{}} n_{\rm eq.} -n \right] + \left[ \vphantom{\frac{}{}} \Thetatreq -\Thetatr \right] \right\},
\end{eqnarray}
where $n_{\rm eq.}$ and $n$ denote the particle number density in and out of equilibrium, respectively. Note that  Eq.~(\ref{UmuUnu_1}) reduces to Eq.~(\ref{bulk_2}) in the limit $m\to 0$.  

\subsubsection{Zeroth moment equation}

The third option we consider is to use the zeroth moment of the Boltzmann equation instead of the second moment. In this case, one obtains
\begin{eqnarray}
D n + n \theta = \frac{1}{\taueq} \left(
n_{\rm eq.} - n \vphantom{\frac{}{}} \right).
\label{zeroth_moment_1}
\end{eqnarray}
It is interesting to observe that Eq.~(\ref{UmuUnu_1}) is the sum of Eq.~(\ref{bulk_2}) and Eq.~(\ref{zeroth_moment_1}) multiplied first by $m^2$. We note, that in general, any linear combination of Eqs.~(\ref{bulk_2}) and~(\ref{zeroth_moment_1}) is acceptable as a scalar equation describing the bulk effects, not only Eq.~(\ref{UmuUnu_1}).

Before proceeding further, we would like to discuss the independent degrees of freedom present in our approach. Since all physical variables are expressed by the non-equilibrium distribution function (\ref{a-Hydro_distribution_0}), they should depend on: the momentum scale parameter $\lambda$, three independent components of the flow vector $U$, five independent components of the tensor $\xi^{\mu\nu}$, and the parameter $\phi$. These are altogether ten independent variables. Thus, in the (3+1)D case we need ten dynamic equations, which can be taken as Eqs.~(\ref{energy_conservation}), (\ref{momentum_conservation}), (\ref{Shear_1}) and one equation out of Eqs.~(\ref{bulk_2}), (\ref{UmuUnu_1}), and (\ref{zeroth_moment_1}). 

We note that the effective temperature $T$ is always calculated from the Landau matching (\ref{Landau_matching}). If we want to treat $T$ as an independent variable, we have to consider the Landau matching as an additional constraint needed to close the system of equations.

\section{Implementation of the Bjorken scenario}
\label{sect:0+1}

\subsection{Local orthonormal basis for the Bjorken flow}

In the Bjorken flow limit, the symmetry of the evolution constrains and greatly simplifies the equations of motion. Because of boost invariance in the longitudinal direction and rotational invariance in the transverse plane, the flow four-velocity $U^\mu$ reads
\begin{equation}\label{U}
 U^{\mu}=(\cosh \eta,0,0,\sinh \eta), \qquad \qquad \eta = \frac{1}{2}\ln\!\left( \frac{t+z}{t-z} \right),
\end{equation}
where $\eta$ is the space-time rapidity. Each scalar function depends in this case only on the longitudinal proper time
\begin{equation}
 \tau = \sqrt{t^2 - z^2}.
\end{equation}
In the local rest frame there is only one privileged space direction, along the beam axis, namely
\begin{equation}\label{Z}
 Z^{\mu}=(\sinh\eta, 0, 0, \cosh\eta), \qquad Z\cdot Z = -1, \qquad Z\cdot U = 0.
\end{equation}
In order to have a complete orthonormal basis, we can choose the other two four-vectors $X$ and $Y$ in such a way that they span the transverse plane and are orthogonal to the $U$ and $Z$ vectors, namely
\begin{equation}
 X\cdot X = -1, \quad Y\cdot Y = -1, \quad X\cdot U = 0, \quad X\cdot Z = 0, \quad Y\cdot U = 0, \quad Y\cdot Z = 0. 
\end{equation}
They can be simply the directions of the $x$ and $y$ coordinates in the lab frame. Therefore, seen from the local rest frame (LRF), they read
\begin{equation}
 U^{\mu} = (1,0,0,0), \quad X^{\mu} = (0,1,0,0), \quad Y^{\mu} = (0,0,1,0), \quad Z^{\mu} = (0,0,0,1). \quad
\end{equation}
The expansion scalar $\theta$ in the (0+1)D case reads
\begin{equation}\label{theta}
\theta = \Delta^{\mu\nu}\partial_\mu U_\nu = \frac{1}{\tau}.
\end{equation}
The tensor $\sigma^{\mu\nu}$ is also relatively simple when expressed in terms of the vectors $X, Y$, and $Z$
\begin{equation}\label{0+1_shear_stress}
 \sigma^{\mu\nu}(x)  = \nabla^{\langle\mu}U^{\nu\rangle} = \sigmatrans(\tau) \left( \vphantom{\frac{}{}} X^\mu X^\nu + Y^{\mu} Y^{\nu} \right) 
 + \sigmalong(\tau) \,  Z^\mu Z^\nu.
\end{equation}
Here the functions $ \sigmatrans$ and $\sigmalong$ are given by 
\begin{equation}\label{sigmas}
 \sigmatrans = \frac{1}{3\tau}, \qquad \qquad \sigmalong = -\frac{2}{3\tau}.
\end{equation}
The condition $2 \sigmatrans + \sigmalong = 0$ reflects the fact that $ \sigma^\mu_{\,\,\mu}=0$.

\subsection{Phase-space distribution function for the (0+1)D case}

In the (0+1)D case,  the anisotropy tensor $\xi^{\mu\nu}$ in~(\ref{decomposition}) reduces to the form
\begin{equation}
 \xi^{\mu\nu} (x) = \xi_\perp(\tau) \left(\vphantom{\frac{}{}} X^\mu X^\nu +Y^\mu Y^\nu \right) + \xi_\|(\tau) Z^\mu Z^\nu,
\end{equation}
which is analogous to (\ref{0+1_shear_stress}). Since $\xi^{\mu\nu}$ is also traceless,  the last equation reads
\begin{equation}
 \xi^{\mu\nu} = -\frac{1}{2}\xi \left(\vphantom{\frac{}{}} X^\mu X^\nu +Y^\mu Y^\nu \right) + \xi\, Z^\mu Z^\nu,
 \label{ximunu}
\end{equation}
where we use the identification $\xi=\xi_\|$. Thus, the distribution function in the local reference frame takes the form
\begin{equation}
 f = \tildeN \exp\!\left[ -\frac{1}{\lambda} \sqrt{  m^2 + \left(1+ \phi -\frac{1}{2}\xi \right)p_\perp^2 + \left(\vphantom{\frac{}{}} 1 + \phi +\xi \right)p_\|^2 } \right].
 \label{f01D}
\end{equation}
The form (\ref{f01D}) suggests that it is useful to introduce the two new $\alpha$ variables defined as
\begin{equation}
 \alphatrans = \frac{1}{\sqrt{  1+\phi-\xi/2  }}, \qquad \qquad \alphalong = \frac{1}{\sqrt{\vphantom{\frac{}{}}  1+\phi +\xi   }},
\end{equation}
and to rewrite the LRF anisotropic distribution function as
\begin{equation}\label{f_0+1}
  f = \tildeN \exp\!\left( -\frac{1}{\lambda} \sqrt{m^2 + \frac{\ptrans^2}{\alphatrans^2} + \frac{\plong^2}{\alphalong^2}} \, \right) .
\end{equation}

\subsection{First-moment tensors}

Using the (0+1)D reduced anisotropic distribution function~(\ref{f_0+1}), the LRF energy density reads 

\begin{equation}\label{energy_density}
 \ped = \tildeN \int d^3{\bf p} \, \sqrt{m^2 +{\bf p}^2} \, \exp\left( -\frac{1}{\lambda} \sqrt{m^2 + \frac{\ptrans^2}{\alphatrans^2} + \frac{\plong^2}{\alphalong^2}} \right) = {\cal \tilde{H}}_3(\alphatrans,\alphalong,\mhat)\lambda^4,
\end{equation}
where $\mhat=m/\lambda$ and  the function ${\cal \tilde{H}}_3$ has been defined in Ref.~\cite{Nopoush:2014pfa}. Because of the symmetry of the system the bulk viscosity $\Pi$, being a scalar, must depend only on the proper time $\tau$, $\Pi = \Pi(\tau)$. On the other hand, if we use the orthonormal basis presented above, we can write the shear tensor $\pi^{\mu\nu}$ in the (0+1)D case as
\begin{equation}
 \pi^{\mu\nu} = \frac{1}{2}\pi_s(\tau)\left(\vphantom{\frac{}{}} X^\mu X^\nu + Y^\mu Y^\nu \right) - \pi_s(\tau) \; Z^\mu Z^\nu.
\end{equation}
It turns out that it is completely described by the space basis $\{X,Y,Z\}$ and the scalar function $\pi_s(\tau)$. However, instead of using $\pi_s$ and $\Pi$, it is sometimes convenient to use the longitudinal pressure $\pres_L$ and the transverse pressure $\pres_T$, 
\begin{equation}\label{PLT}
 \pres_L = Z\cdot T\cdot Z = \preseq +\Pi -\pi_s, \qquad \qquad \pres_T = X\cdot T \cdot X = Y\cdot T\cdot Y = \preseq + \Pi + \frac{1}{2}\pi_s,
\end{equation}
since this results in relatively simple expressions, which do not depend on the effective temperature, namely

\begin{eqnarray}\label{PL}
 \pres_L &=& \tildeN \int d^3{\bf p} \, \frac{\plong^2}{ \sqrt{m^2 +{\bf p}^2}}\,  \exp\left( -\frac{1}{\lambda} \sqrt{m^2 + \frac{\ptrans^2}{\alphatrans^2} + \frac{\plong^2}{\alphalong^2}} \right) = {\cal \tilde{H}}_{3L}(\alphatrans,\alphalong,\mhat)\lambda^4, \\\label{PT}
 \pres_T &=& \frac{\tildeN}{2} \int d^3{\bf p} \, \frac{\ptrans^2}{ \sqrt{m^2 +{\bf p}^2}}\,  \exp\left( -\frac{1}{\lambda} \sqrt{m^2 + \frac{\ptrans^2}{\alphatrans^2} + \frac{\plong^2}{\alphalong^2}} \right) = {\cal \tilde{H}}_{3T}(\alphatrans,\alphalong,\mhat)\lambda^4.
\end{eqnarray}
Here the integrals are taken in the local rest frame, and the functions ${\cal \tilde{H}}_{3L}$ and ${\cal \tilde{H}}_{3T}$ are also defined in Ref.~\cite{Nopoush:2014pfa}.

One can recover the original degrees of freedom $\Pi$ and $\pi_s$ from the longitudinal pressure, the transverse pressure and the hydrostatic pressure in Eq.~(\ref{Peq}) using 

\begin{eqnarray}
 \Pi = \frac{1}{3}\left(\vphantom{\frac{}{}}2\pres_T + \pres_L\right) -\preseq, \qquad \qquad \pi_s = \frac{2}{3} \left(\vphantom{\frac{}{}} \pres_T - \pres_L \right).
\end{eqnarray}
We note that only the integrals in Eqs.~(\ref{energy_density}),~(\ref{PL}), and~(\ref{PT}) depend on the anisotropic distribution function.  The rest of the arguments stem directly from the symmetry of the system, which is completely general in the (0+1)D set up.

\subsection{Second- and zeroth-moment tensors}

Similarly to Eqs.~(\ref{0+1_shear_stress}) and (\ref{ximunu}), the tensor $\Theta^{\mu\nu}$ can be decomposed as
\begin{equation}\label{0+1_Theta}
 \Theta^{\mu\nu} = \Thetatrans(\tau) \left( X^\mu X^\nu + Y^\mu Y^\nu \vphantom{\frac{}{}}\right) + \Thetalong(\tau) Z^\mu Z^\nu,
\end{equation}
and, consequently, we find
\begin{equation}\label{0+1_Theta_traceless}
 \Theta^{\langle\mu\nu\rangle} = \frac{1}{3}\left(\vphantom{\frac{}{}}\Thetatrans -\Thetalong \right) \left(\vphantom{\frac{}{}} X^\mu X^\nu + Y^\mu Y^\nu \right) +\frac{2}{3}\left(\vphantom{\frac{}{}} \Thetalong -\Thetatrans \right) Z^\mu Z^\nu.
\end{equation}
Using Eq.~(\ref{f_0+1}), the functions $\Thetatrans$ and $\Thetalong$ can be written as follows
\begin{eqnarray}\label{Thetatrans}
 \Thetatrans = \alpha \alphatrans^2 (4\pi\tildeN)\lambda^5 \mhat^3K_3(\mhat), \\ \nonumber \\ \label{Thetalong}
 \Thetalong = \alpha \alphalong^2 (4\pi\tildeN)\lambda^5 \mhat^3 K_3(\mhat),
\end{eqnarray}
where $\alpha = \alphatrans^2\alphalong$.

The last integral over the anisotropic distribution function which is necessary is the (non-equilibrium) number density
 \begin{equation}
  n = \alpha n_{\rm eq.}(\lambda),
 \end{equation}
 with $n_{\rm eq.}(\lambda)$ being the local equilibrium number density with the substitution $T \to \lambda$. The equilibrium density reads
 \begin{equation}
  n_{\rm eq.} = (4\pi \tildeN) T^3 \mheq ^2 K_2(\mheq).
 \end{equation}
 %

\section{Reduced evolution equations}
\label{sect:reducedeq}

\subsection{Energy conservation}
\label{sect:shear}

Since we are using the same form of the anisotropic distribution function as that used in Ref.~\cite{Nopoush:2014pfa},  our equations expressing the energy-momentum conservation are the same as in~\cite{Nopoush:2014pfa}. Using Eq.~(\ref{energy_density}), the Landau matching~(\ref{Landau_matching}) can be written as
 \begin{equation}
  {\cal \tilde{H}}_3(\alphatrans,\alphalong,\mhat)\lambda^4 = \pedeq(T, \mheq).
 \end{equation}
 The three-momentum conservation equations in~(\ref{momentum_conservation}) are automatically satisfied by symmetry. Indeed, using Eqs.~(\ref{U}) and~(\ref{Z}) one finds that
\begin{equation}
 D = U \cdot\partial  = \partial_\tau,  \qquad Z \cdot\partial  = \frac{1}{\tau} \partial_\eta,  \qquad DU^\mu = 0, \qquad (Z\cdot\partial) Z^\mu = \frac{1}{\tau} U^\mu, \qquad  \partial\cdot Z = 0.
\end{equation}
Using these relations we check that both the left- and right-hand sides of Eq.~(\ref{momentum_conservation}) vanish. In particular, for the right-hand side of Eq.~(\ref{momentum_conservation}) we obtain
\begin{eqnarray}
 \nabla^\mu \preseq(\tau) +\nabla^\mu \Pi(\tau) +\Delta^\mu_\rho \partial_\sigma \pi^{\rho\sigma} &=& \pi_s(\tau) \, \Delta^\mu_\rho \,\partial_\sigma \left[  \frac{1}{2}\left( \vphantom{\frac{}{}} X^\rho X^\sigma +Y^\rho Y^\sigma \right) - Z^\rho Z^\sigma \right] \\ 
 &=& -\pi_s(\tau)\left[ \vphantom{\frac{}{}} \Delta^\mu_\rho (Z\cdot\partial) Z^\rho +Z^\mu \partial\cdot Z \right] =0.
\end{eqnarray}
In a similar manner, from the energy conservation equation~(\ref{energy_conservation}) one obtains
\begin{equation}
 D\ped = -\frac{\ped +\preseq +\Pi-\pi_s}{\tau} = -\frac{\ped +\pres_L}{\tau}.
 \label{emconsBjorken}
\end{equation}
Here we used the definition of the longitudinal pressure~(\ref{PLT}). Using Eqs.~(\ref{energy_density}),~(\ref{PL}), and~(\ref{PT}) the last equation can be rewritten as
\begin{equation}
 \left( \vphantom{\frac{}{}} 4{\cal \tilde{H}}_3(\alphatrans,\alphalong,\mhat) -\tilde{\Omega}_m(\alphatrans,\alphalong,\mhat) \right)D\ln\lambda + 2 \tilde{\Omega}_T(\alphatrans,\alphalong,\mhat) D\ln\alphatrans +\tilde{\Omega}_L(\alphatrans,\alphalong,\mhat) \left(\frac{1}{\tau} + D\ln\alphalong \right) =0.
\end{equation}
For the explicit expressions of the functions $\tilde{\Omega}_m$, $\tilde{\Omega}_T$, and $\tilde{\Omega}_L$ see Ref.~\cite{Nopoush:2014pfa}.

\subsection{Shear viscosity equations}
\label{sect:shear}

The equations describing the shear pressure corrections are the same for the three different options for the bulk viscous pressure evolution equations, and they can be rewritten as
\begin{equation}\label{shear_1}
 \Delta^{\mu\nu}_{\alpha\beta} D\Theta^{\alpha\beta} + \frac{5}{3}\theta \, \Theta^{\langle\mu\nu\rangle} +  2\Theta^{\langle\mu}_\lambda  \sigma^{\nu\rangle\lambda} -2\Theta^{\langle\mu}_\lambda \omega^{\nu\rangle\lambda} = -\frac{1}{\taueq} \Theta^{\langle\mu\nu\rangle}.
\end{equation}
Since the convective derivatives of $X$, $Y$, and $Z$ either vanish or are proportional to $U$, the first term of the last equation reads
\begin{equation}
  \Delta^\mu_\alpha\Delta^\nu_\beta D\Theta^{\alpha\beta} -\frac{1}{3}\Delta^{\mu\nu} \Delta_{\alpha\beta} D \Theta^{\alpha\beta} = \frac{1}{3}\left(\vphantom{\frac{}{}} D\Thetatrans -D\Thetalong  \right) \left( \vphantom{\frac{}{}} X^\mu X^\nu + Y^\mu Y^\nu \right) + \frac{2}{3}\left( \vphantom{\frac{}{}} D\Thetalong -D\Thetatrans \right) Z^\mu Z^\nu.
\end{equation}
The contraction of $\Theta$ with the expansion tensor $\sigma$ reads
\begin{equation}
  \Theta^{\langle\mu}_\lambda  \sigma^{\nu\rangle\lambda} = -\frac{1}{3}\left( \vphantom{\frac{}{}} \Thetatrans\sigma_\perp - \Thetalong \sigma_\| \right) \left(\vphantom{\frac{}{}} X^\mu X^\nu + Y^\mu Y^\nu \right) - \frac{2}{3}\left( \frac{}{} \Thetalong\sigma_\| -\Thetatrans\sigma_\perp \right) Z^\mu Z^\nu,
\end{equation}
where we used the (0+1)D form of the expansion tensor~(\ref{0+1_shear_stress}). Since the vorticity $\omega^{\mu\nu}$ vanishes in the case of (0+1)D expansion, collecting all of the terms together in Eq.~(\ref{shear_1}) one finds
\begin{eqnarray}\nonumber
 && \frac{1}{3}\left[ D\Thetatrans -D\Thetalong + \frac{5}{3}\theta \left( \vphantom{\frac{}{}}\Thetatrans -\Thetalong \right) -2 \left( \vphantom{\vphantom{\frac{}{}}} \Thetatrans\sigma_\perp - \Thetalong \sigma_\| \right) +\frac{1}{\taueq}\left( \vphantom{\frac{}{}}\Thetatrans -\Thetalong \right) \right] \left(\vphantom{\frac{}{}} X^\mu X^\nu + Y^\mu Y^\nu \right) + \\
 && \qquad -\frac{2}{3}\left[ D\Thetatrans -D\Thetalong + \frac{5}{3}\theta \left( \vphantom{\frac{}{}}\Thetatrans -\Thetalong \right) -2 \left(\vphantom{\frac{}{}} \Thetatrans\sigma_\perp - \Thetalong \sigma_\| \right) +\frac{1}{\taueq}\left(\vphantom{\frac{}{}}\Thetatrans -\Thetalong \right) \right] Z^\mu Z^\nu =0.
\end{eqnarray}
The equations for the shear pressure corrections are then straightforwardly fulfilled if
\begin{equation}\label{shear_0+1_1}
 D\Thetatrans -D\Thetalong + \frac{5}{3}\theta \left(\vphantom{\frac{}{}}\Thetatrans -\Thetalong \right) -2 \left( \vphantom{\frac{}{}} \Thetatrans\sigma_\perp - \Thetalong \sigma_\| \right) +\frac{1}{\taueq}\left( \vphantom{\frac{}{}}\Thetatrans -\Thetalong \right) = 0.
\end{equation}
We note that the number of independent equations in the (0+1)D case is reduced to one.  This is not surprising, since the number of equations matches the number of independent parameters defining the anisotropy tensor $\xi^{\mu \nu}$. To simplify the last expression we can use the explicit expressions for the four velocity gradients~(\ref{sigmas}) 
\begin{equation}\label{shear_0+1_2}
 D\left( \vphantom{\frac{}{}} \Thetatrans - \Thetalong \right) +\frac{1}{\tau}\left( \vphantom{\frac{}{}} \Thetatrans - \Thetalong \right) -\frac{2}{\tau}\Thetalong + \frac{1}{\taueq}\left(\vphantom{\frac{}{}} \Thetatrans - \Thetalong \right) = 0.
\end{equation}
Using the expressions (\ref{Thetatrans}) and (\ref{Thetalong}), and dividing (\ref{shear_0+1_2}) by the {\it positive} quantity $\alpha  (4\pi\tildeN)\lambda^2 m^3K_3(m/\lambda)$, Eq.~(\ref{shear_0+1_2}) is put into the form 
\begin{eqnarray}\label{shear_0+1_3}
 && \left( \vphantom{\frac{}{}} 4\alphatrans^2 -2\alphalong^2 \right)D\ln\alphatrans + \left(\vphantom{\frac{}{}} \alphatrans^2 - 3\alphalong^2 \right)D\ln\alphalong + \left( \vphantom{\frac{}{}} \alphatrans^2 - \alphalong^2 \right) \left[5 +\mhat  \frac{K_2(\mhat)}{K_3(\mhat)} \right] D\ln\lambda + \\ \nonumber
 && \qquad +  \left( \vphantom{\frac{}{}} \alphatrans^2 - \alphalong^2 \right)\left( \frac{1}{\tau} + \frac{1}{\taueq} \right ) - \frac{2\alphalong^2}{\tau} = 0.
\end{eqnarray}

\subsection{Bulk viscosity equations}
\label{sect:bulk}

\subsubsection{Trace equation}

Using the decomposition~(\ref{0+1_Theta}) and~(\ref{0+1_shear_stress}) the trace of the second moment of the Boltzmann equation reads

\begin{equation}
 -D\left(\vphantom{\frac{}{}} 2\Thetatrans +\Thetalong \right) - \frac{5}{3} \, \theta \, \left( \vphantom{\frac{}{}} 2\Thetatrans +\Thetalong \right) + 2 \left(\vphantom{\frac{}{}} 2\Thetatrans\sigma_\perp +\Thetalong\sigma_\| \right) = -\frac{1}{\taueq}\left[\vphantom{\frac{}{}} 3 \Thetaeq - \left( \vphantom{\frac{}{}} 2\Thetatrans +\Thetalong \right) \right].
 \label{tracebulk}
\end{equation}
As was done for the shear pressure corrections equations, it is possible to use the Eq.~(\ref{Thetatrans}),~(\ref{Thetalong}),~(\ref{theta}), and~(\ref{sigmas}) to further simplify the Eq. (\ref{tracebulk}). Dividing by the positive function $\alpha  (4\pi\tildeN)\lambda^2 m^3K_3(m/\lambda)$ one obtains
\begin{eqnarray}\label{traceeq}
 && -\left( \vphantom{\frac{}{}} 8\alphatrans^2 +2\alphalong^2 \right)D\ln\alphatrans - \left( \vphantom{\frac{}{}} 2\alphatrans^2 + 3\alphalong^2 \right)D\ln\alphalong - \left(  \vphantom{\frac{}{}} 2\alphatrans^2 + \alphalong^2 \right) \left[5 +\mhat  \frac{K_2(\mhat)}{K_3(\mhat)} \right] D\ln\lambda + \\ \nonumber\label{bulk_0+1}
 && \qquad  - \left(  \vphantom{\frac{}{}} 2\alphatrans^2 + \alphalong^2 \right)\left( \frac{1}{\tau} + \frac{1}{\taueq} \right ) - \frac{2\alphalong^2}{\tau} \; + \;  \frac{3}{\taueq}\, \frac{T^2}{\alpha\lambda^2} \, \frac{K_3(\mheq)}{K_3(\mhat)}= 0.
\end{eqnarray}

\subsubsection{U projection}

In the case where the $U$ projection is applied we use the equation
\begin{eqnarray}\label{uueq}
 && \mhat K_2(\mhat) \left\{ D\ln\alpha  + \left[ 3 +\mhat\frac{K_1(\mhat)}{K_2(\mhat)} \right] D\ln\lambda +\frac{1}{\tau} + \frac{1}{\taueq}\left[ 1 - \frac{T}{\alpha\lambda} \, \frac{K_2(\mheq)}{K_2(\mhat)} \right] \right\} \\ \nonumber
 && +K_3(\mhat) \left\{ -\left( \vphantom{\frac{}{}} 8\alphatrans^2 +2\alphalong^2 \right)D\ln\alphatrans - \left(  \vphantom{\frac{}{}} 2\alphatrans^2 + 3\alphalong^2 \right)D\ln\alphalong - \left(  \vphantom{\frac{}{}} 2\alphatrans^2 + \alphalong^2 \right) \left[5 +\mhat  \frac{K_2(\mhat)}{K_3(\mhat)} \right] D\ln\lambda + \right. \\
 && \left. \qquad  - \left(  \vphantom{\frac{}{}} 2\alphatrans^2 + \alphalong^2 \right)\left( \frac{1}{\tau} + \frac{1}{\taueq} \right ) - \frac{2\alphalong^2}{\tau} \; + \;  \frac{3}{\taueq}\, \frac{T^2}{\alpha\lambda^2} \, \frac{K_3(\mheq)}{K_3(\mhat)}\right\}= 0.\nonumber
\end{eqnarray}

\subsubsection{Zeroth moment equation} 

The zeroth moment of the Boltzmann equation was previously obtained in Ref.~\cite{Nopoush:2014pfa} using the same form~(\ref{f_0+1}) for the anisotropic distribution function, with the result being

\begin{eqnarray}\label{zerothmomeq}
 D\ln\alpha  + \left[ 3 +\mhat\frac{K_1(\mhat)}{K_2(\mhat)} \right] D\ln\lambda +\frac{1}{\tau} + \frac{1}{\taueq}\left[ 1 - \frac{T}{\alpha\lambda} \, \frac{K_2(\mheq)}{K_2(\mhat)} \right]= 0.
\end{eqnarray}

\begin{figure}[t]
\begin{center}
\includegraphics[width=1.0\columnwidth]{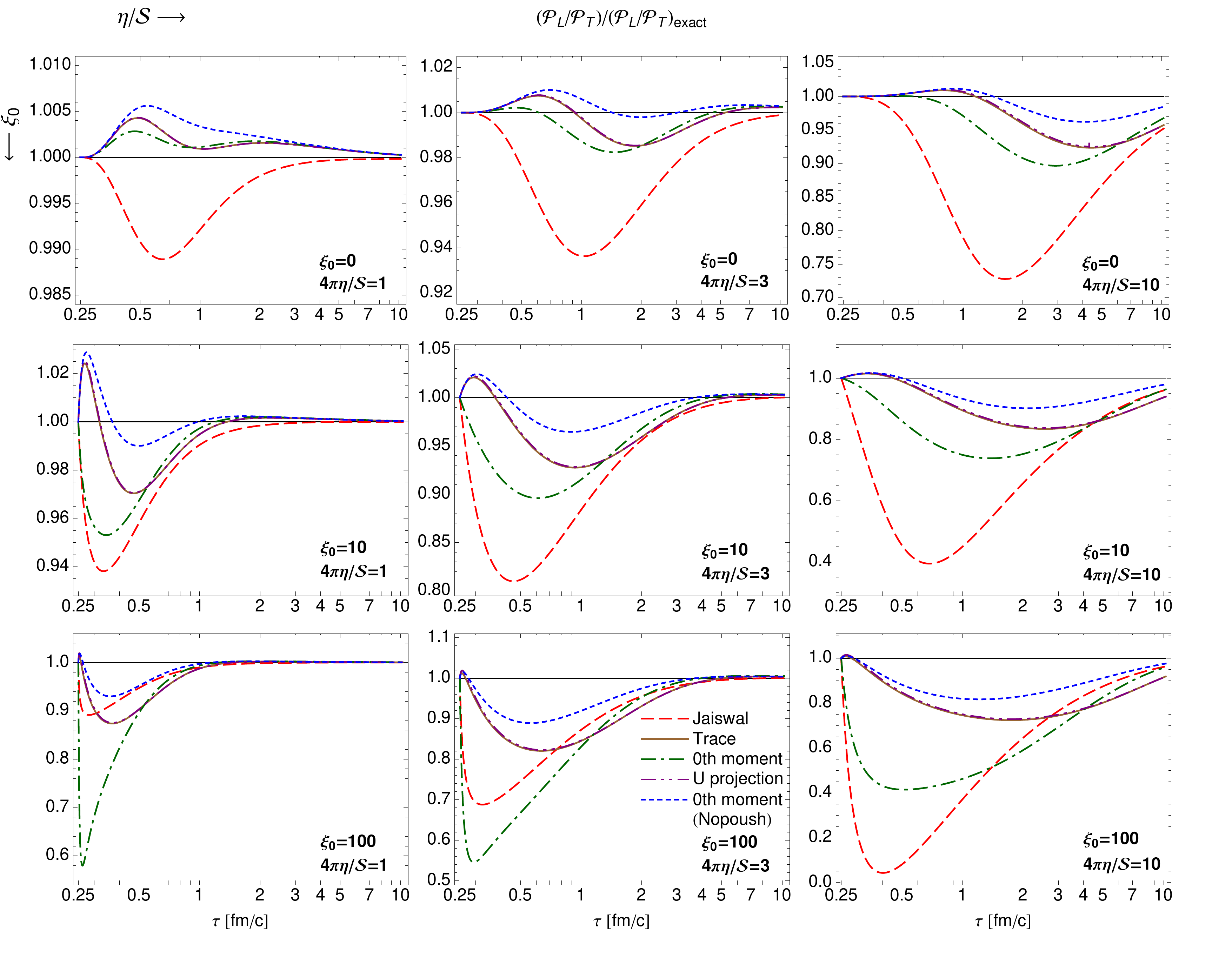}
\end{center}
\caption{(Color online) Ratio of the longitudinal and transverse pressures plotted as a function of the longitudinal proper time. The results of various hydrodynamic calculations are normalized to the exact result obtained from the kinetic theory. Details of the figure are explained in the text. }
\label{fig:NONLinearPL2PT}
\end{figure}

\begin{figure}[t]
\begin{center}
\includegraphics[width=1.\columnwidth]{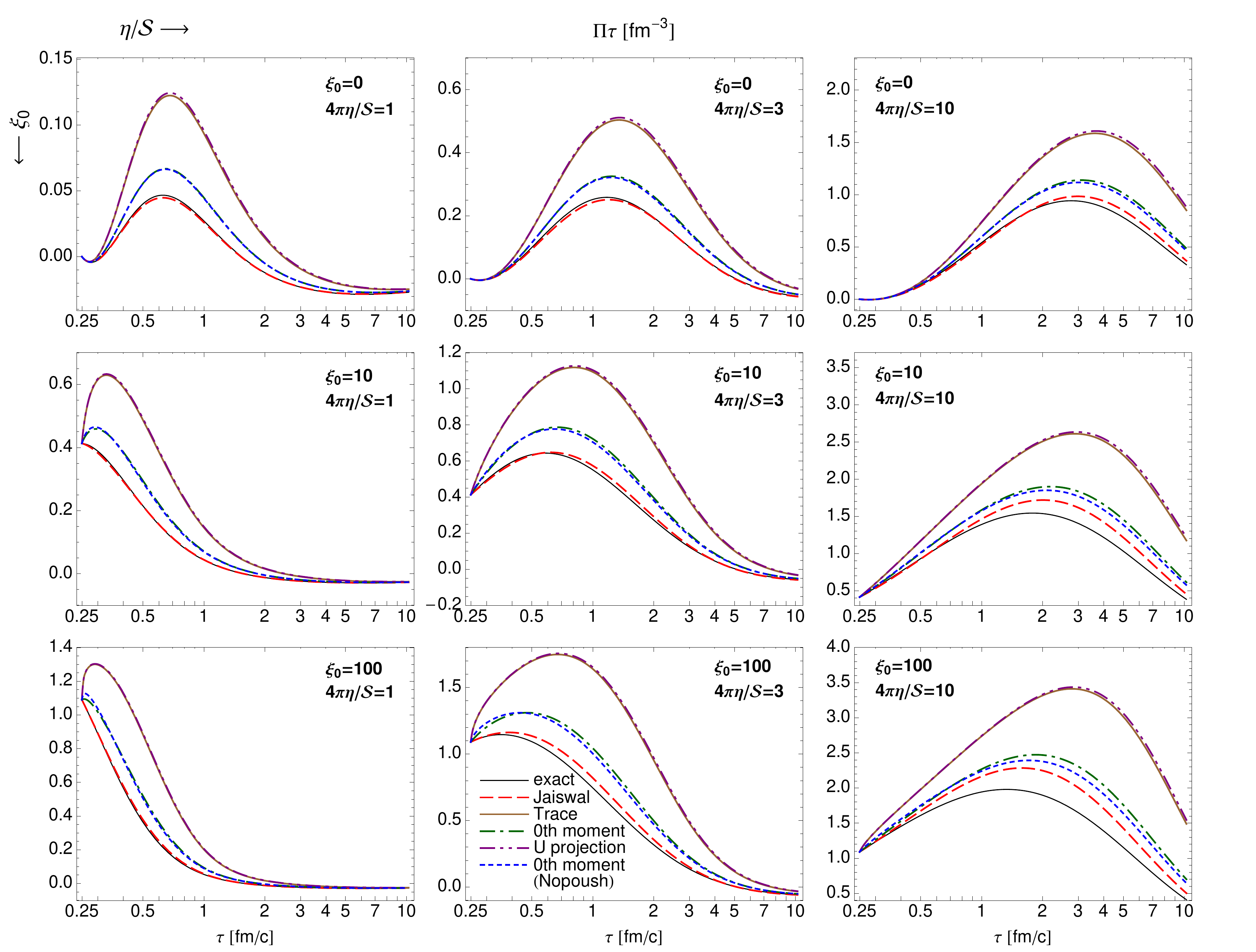}
\end{center}
\caption{(Color online)  Proper time dependence of the bulk viscous pressure multiplied by the proper time. The hydrodynamic calculations are compared to the exact solution of the Boltzmann equation (thin solid black line).  The remaining notation is the same as in Fig.~\ref{fig:NONLinearPL2PT}.}%
\label{fig:NONLinearBULK}
\end{figure}

\begin{figure}
\begin{minipage}[t]{0.47 \linewidth}
\centering
\includegraphics[angle=0,width=0.84 \textwidth]{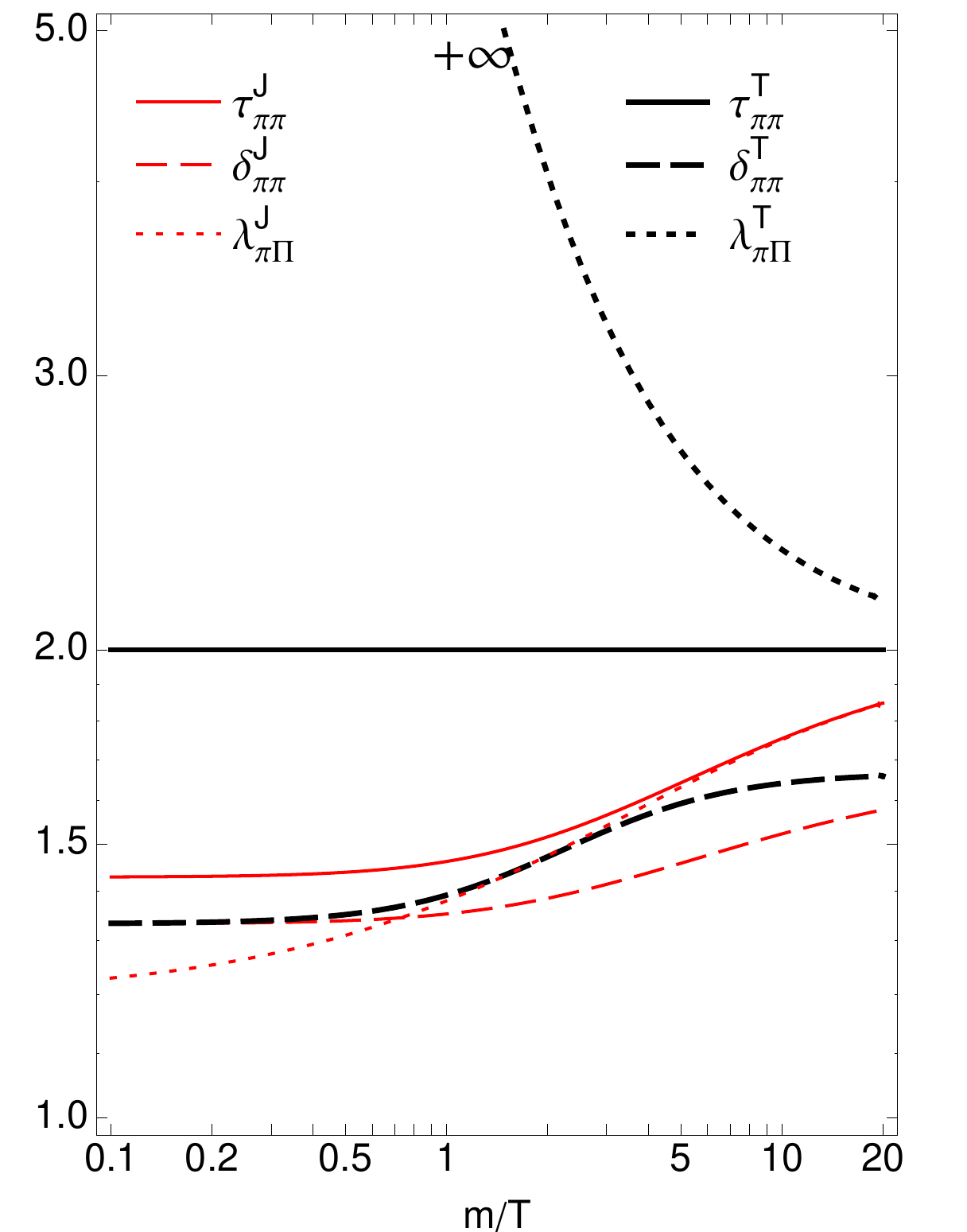}
\end{minipage}
\begin{minipage}[t]{0.47\linewidth}
\centering
\includegraphics[angle=0,width=0.84 \textwidth]{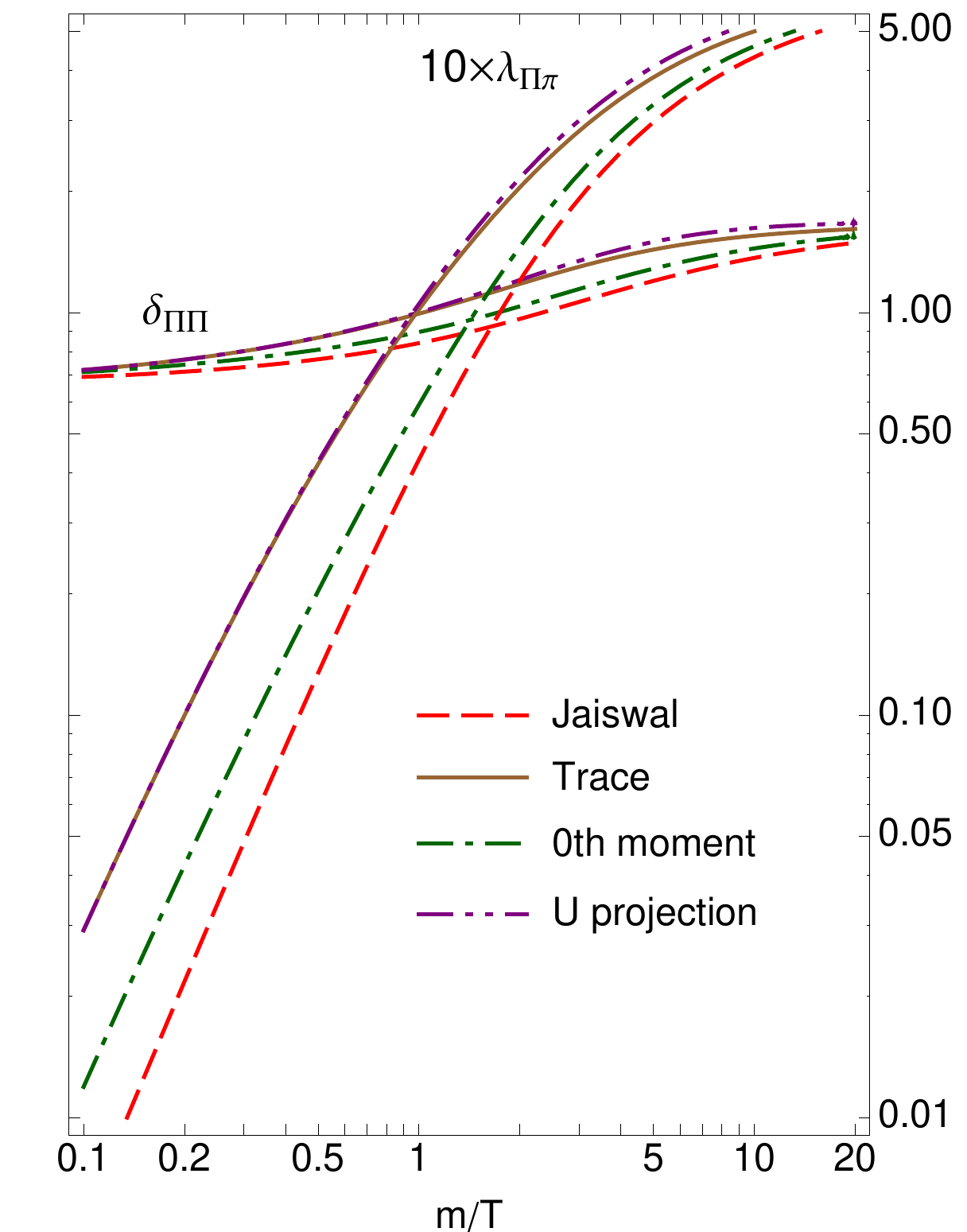}
\end{minipage}
\caption{(Color online) Comparison of the transport coefficients. Left panel: Transport coefficients $\tau_{\pi\pi}, \delta_{\pi\pi}$, and $\lambda_{\pi\Pi}$ from Ref.~\cite{Jaiswal:2013npa} (red lines) and \cite{Tinti:2014cpa} (black lines).  Right panel: Transport coefficients $\lambda_{\Pi\pi}$ (multiplied by 10) and  $\delta_{\Pi\Pi}$ from Ref.~\cite{Jaiswal:2013npa} (dashed lines) and from \cite{Tinti:2014cpa} (three different formulations shown as solid, dashed-dotted and dashed-double-dotted lines). All transport coefficients are plotted as functions of the ratio $m/T$.}
\label{fig:transport_coefficients1}
\end{figure}

\section{Numerical results}
\label{sect:num}

In this section we present our numerical results.  We first present comparisons of solution of the full non-perturbative equations of motion and then make comparisons using the linearized equations obtained in the near-equilibrium limit.  In all figures, we take the initial proper time to be $\tau_0=$ 0.25 fm/c and, when a fixed mass is required, we take $m = 300$ MeV.  We evolve the equations subject to the assumption of fixed ratio of the viscosity, $\eta$, to equilibrium entropy density, ${\cal S}_{\rm eq.}$.  The equilibrium entropy density ${\cal S}_{\rm eq.}$ at a given effective temperature $T$ is obtained using the equilibrium relation 
\begin{eqnarray}
{\cal S}_{\rm eq.}(T) = \frac{\pedeq(T)+\preseq(T)}{T},
\label{calS}
\end{eqnarray} 
while, for the shear viscosity, we use the formula~\cite{Anderson:1974,Czyz:1986mr}
\begin{equation}
\eta(T) = \frac{\taueq(T) {\cal P}_{\rm eq}(T)}{15} \, \mheq^3\left[ \frac{3}{\mheq^2}\frac{K_3}{K_2}  -\frac{1}{\mheq}+\frac{K_1}{K_2}-\frac{K_{i,1}}{K_2} \right] \, ,
\label{etaAW}
\end{equation}
which is valid for a relativistic classical gas of massive particles. Here all the functions should be evaluated at $\mheq$, $K_n$~are modified Bessel functions, and $2 K_{i,1} = \pi\left[1 - \mheq K_0(\mheq) L_{-1}(\mheq) - \mheq K_1(\mheq) L_{0}(\mheq) \right]$ where $L_i$ is a modified Struve function \cite{Florkowski:2014sfa}. From Eqs.~(\ref{calS}) and (\ref{etaAW}) we find the relaxation time $\taueq$ as a function of the effective temperature. The latter can be used directly in both hydrodynamic and kinetic calculations. In the remainder of this section, we will drop the `$\rm eq.$' subscript from the entropy with the understanding that the entropy appearing in the ratio $\eta/{\cal S}$ is the equilibrium entropy density.

\subsection{Non-perturbative case}
\label{sect:num1}

In Fig.~\ref{fig:NONLinearPL2PT} we show our numerical results for the ratio of the longitudinal and transverse pressures, ${\cal P}_L/{\cal P}_T$, plotted as a function of the longitudinal proper time $\tau$. The behavior of this ratio is determined mainly by effects connected with the shear viscosity.   The ratio determined from the various hydrodynamic calculations is normalized to the exact result obtained from kinetic theory.\footnote{For the detailed discussion of the method used for this purpose we refer the reader to \cite{Florkowski:2014sfa}.} The upper, middle, and lower panels of Fig.~\ref{fig:NONLinearPL2PT} are the results obtained with the initial anisotropy parameter $\xi_0 = \xi(\tau_0)=0$, 10, and 100, respectively. Similarly, the three columns are the results obtained using different values of the ratio of the viscosity to the entropy density; $4\pi\eta/{\cal S} = 1$ in the first column, $4\pi\eta/{\cal S}  = 3$ in the second column, and $4\pi\eta/{\cal S} = 10$ in the third column.  

The solid (brown), dashed-dotted (green), and dashed-double-dotted (violet) lines in Fig.~\ref{fig:NONLinearPL2PT} show the results obtained using leading-order anisotropic hydrodynamics with the use of Eqs.~(\ref{traceeq}), (\ref{uueq}), and (\ref{zerothmomeq}), respectively. The long-dashed (red) line represents the results obtained with the recent second-order hydrodynamic formulation by Jaiswal \cite{Jaiswal:2013npa}, while the short-dashed (blue) lines show the results obtained within an earlier formulation of the leading order anisotropic hydrodynamics, designed for the (1+1)D massless case in \cite{Tinti:2013vba} and generalized to the massive case by Nopoush et al. in Ref.~\cite{Nopoush:2014pfa}. The results presented in Fig.~\ref{fig:NONLinearPL2PT} indicate that, as long as the initial anisotropy is relatively small, $\xi_0 \leq 10$, all versions of leading-order anisotropic hydrodynamics give better agreement with the exact result than second-order hydrodynamics equations of Ref.~\cite{Jaiswal:2013npa}. For larger initial anisotropies the versions of anisotropic hydrodynamics based on the trace or on the $U$ projection of the second moment, see Eqs.~(\ref{traceeq}) and (\ref{uueq}), seem to be preferred.\footnote{One should note here that, although the results obtained within the (1+1)D formulation obtained in Ref.~\cite{Nopoush:2014pfa} work better than all formulations obtained in \cite{Tinti:2014cpa} in the entire (${\bar\eta}, \xi_0$) parameter space, in their present form they cannot be used in the most realistic full (3+1)D evolution.}

In Fig.~\ref{fig:NONLinearBULK} we show our numerical results for the proper time dependence of the bulk viscous pressure multiplied by the proper time. The notation is the same as in Fig.~\ref{fig:NONLinearPL2PT}.  In this case, the best description is given by the second-order hydrodynamics equations of Ref.~\cite{Jaiswal:2013npa}, but anisotropic hydrodynamics using the zeroth moment, see Eq.~(\ref{zerothmomeq}), also gives satisfactory results. It is interesting to observe that very good agreement with the exact results is always obtained within the leading order (1+1)D formulation defined in Refs.~\cite{Tinti:2013vba,Nopoush:2014pfa} (short-dashed blue lines).

\begin{figure}
\begin{minipage}[t]{0.47 \linewidth}
\centering
\includegraphics[angle=0,width=0.84 \textwidth]{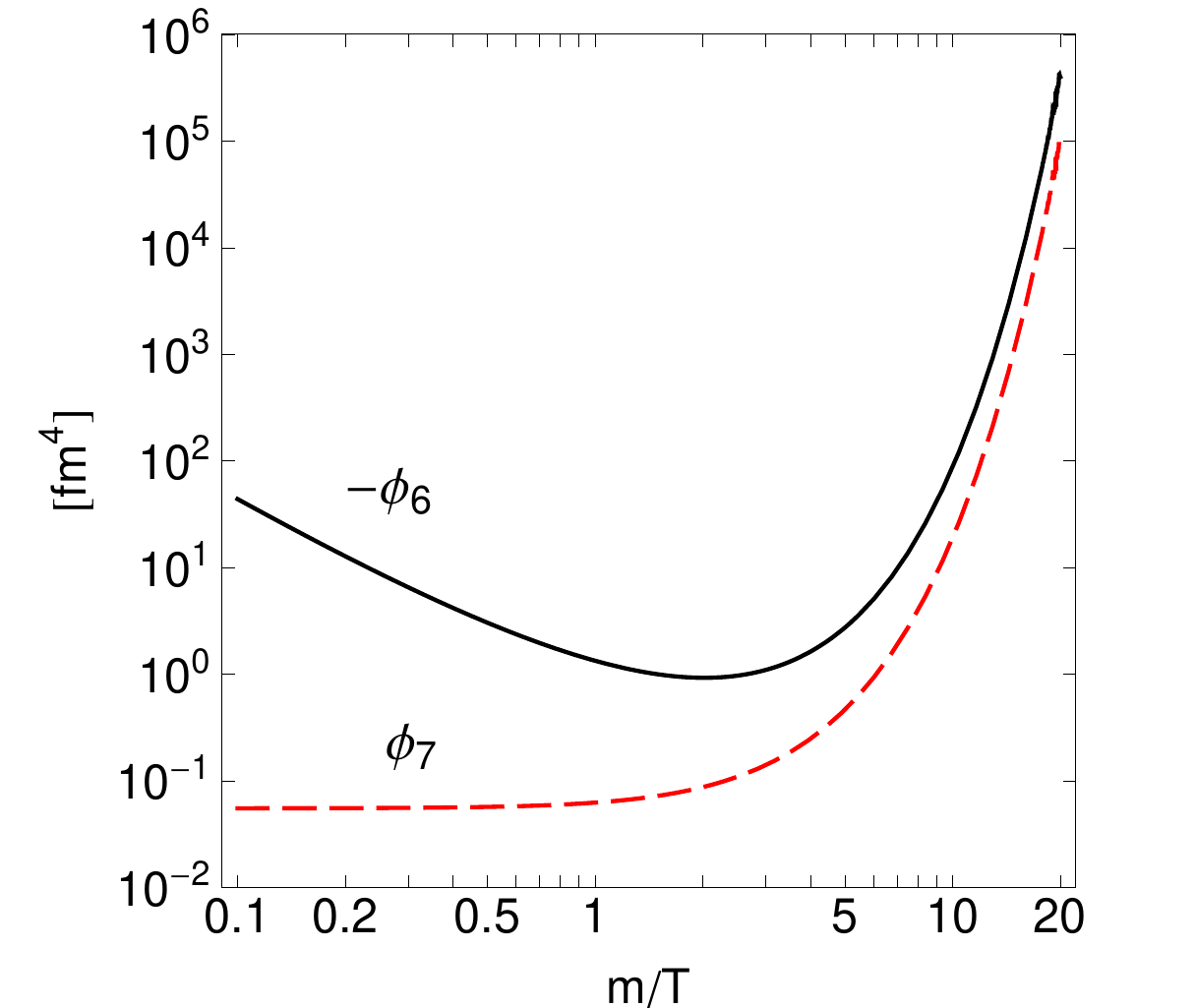}
\end{minipage}
\begin{minipage}[t]{0.47\linewidth}
\centering
\includegraphics[angle=0,width=0.84 \textwidth]{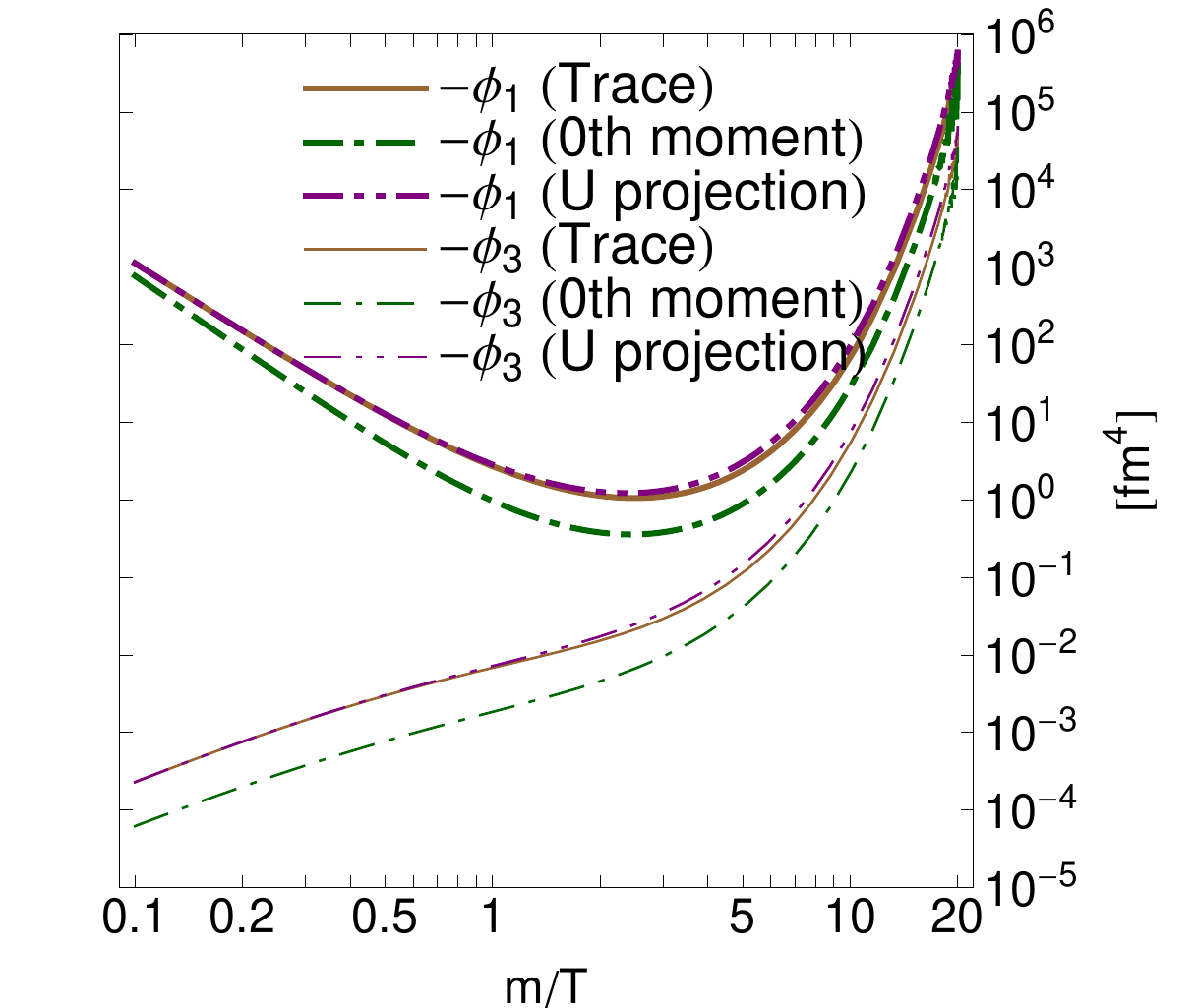}
\end{minipage}
\caption{(Color online) Left panel: Transport coefficients $\phi_6$ (solid black line) and $\phi_7$ (dashed red line) appearing in the evolution equation for the shear tensor in the linearised version ($\phi_6$ and $\phi_7$ are independent of the prescription chosen to describe the bulk pressure).  Right panel: Transport coefficients $\phi_1$ (thick lines) and $\phi_3$ (thin lines) for three prescriptions of the bulk pressure evolution defined by Eqs.~~(\ref{traceeq}), (\ref{uueq}) and (\ref{zerothmomeq}).} 
\label{fig:transport_coefficients2}
\end{figure}

\begin{figure}[t]
\begin{center}
\includegraphics[width=1.0\columnwidth]{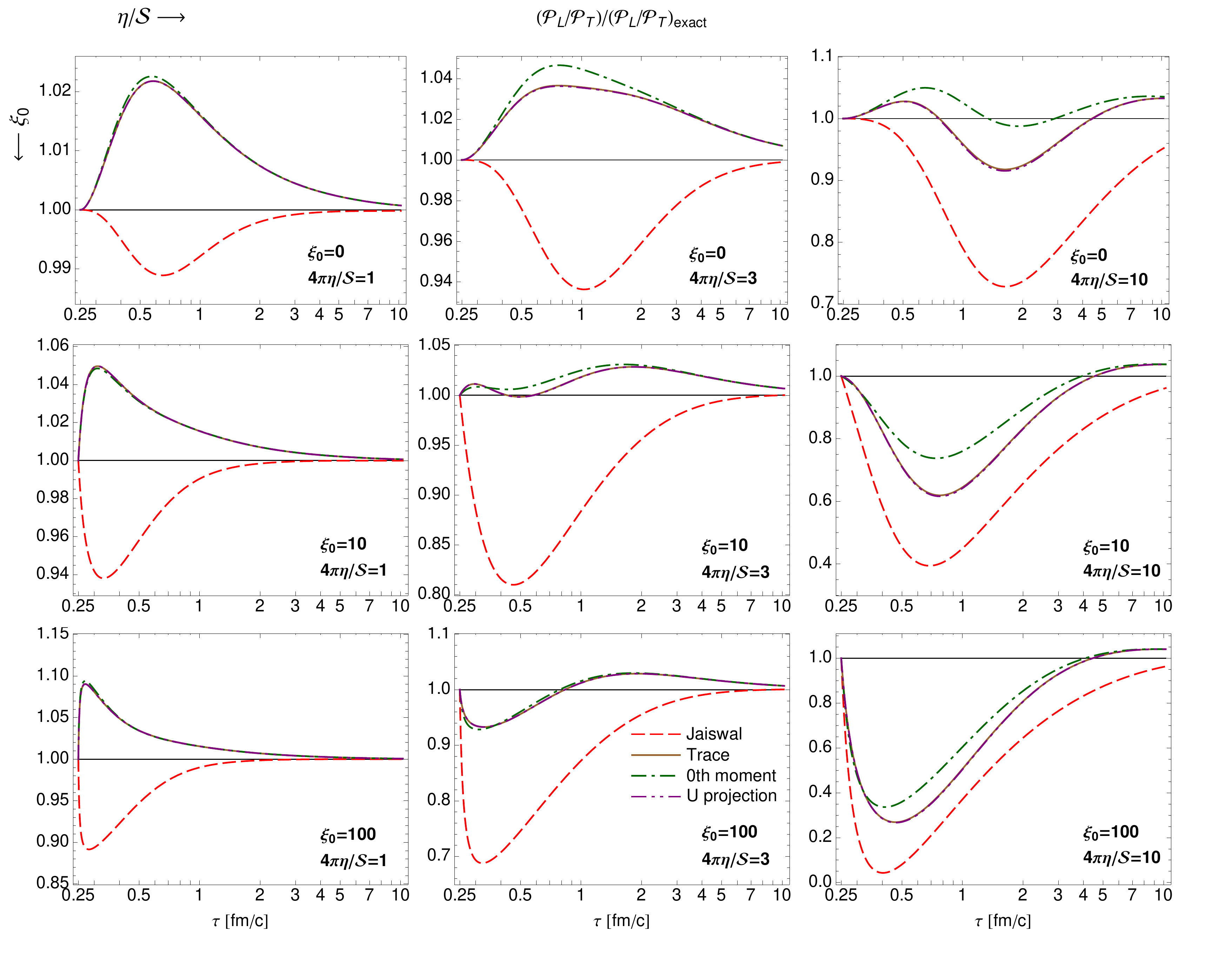}
\end{center}
\caption{(Color online)  Same as Fig.~\ref{fig:NONLinearPL2PT} but for the close-to-equilibrium limit. } 
\label{fig:LinearPL2PT}
\end{figure}

\begin{figure}[t]
\begin{center}
\includegraphics[width=1.\columnwidth]{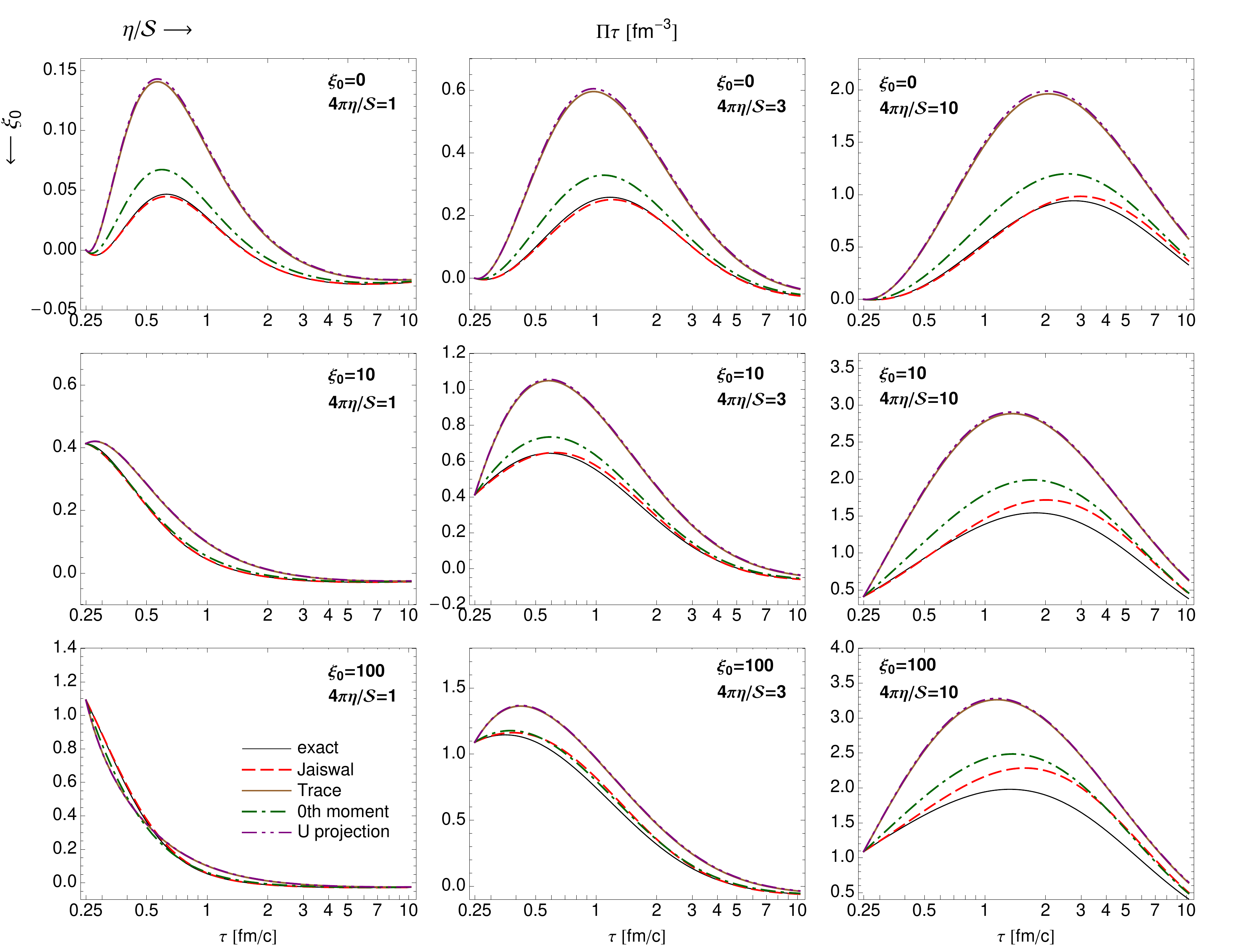}
\end{center}
\caption{(Color online)  Same as Fig.~\ref{fig:NONLinearBULK} but for the close-to-equilibrium limit.}
\label{fig:LinearBULK}
\end{figure}

\subsection{Close-to-equilibrium limit}
\label{sect:num2}

Despite the good agreement with the exact solutions seen in the last section, one may ask how much this agreement depends on the non-perturbative treatment of the large pressure anisotropy. It is possible that the close-to-equilibrium behavior is not well reproduced, but the non-linear treatment of the anisotropy compensates for it.  In such a case, the set of equations would not be reliable for small pressure corrections. In heavy-ion collisions, however, the pressure corrections in the transverse plane are relatively small, therefore, the proposed equations should be reliable even for small pressure corrections.

It has been shown in Ref.~\cite{Tinti:2014cpa} that in the close-to-equilibrium limit all prescriptions for the (3+1)D anisotropic hydrodynamics dynamical equations give a structure that is of the same form as obtained using second order viscous hydrodynamics \cite{Denicol:2014loa}. One can easily show that, for the Bjorken scenario considered herein, Eqs.~(42) and (43) from Ref.~\cite{Tinti:2014cpa} reduce to 
\begin{align}
D\Pi + \frac{\Pi}{\tau_\Pi} &= -\frac{\zeta}{\tau_\Pi\tau} - \delta_{\Pi\Pi}\frac{\Pi}{\tau}
+\lambda_{\Pi\pi}\frac{\pi_s}{\tau} + \frac{\phi_1}{\tau_\Pi}  \Pi^2 + \frac{\phi_3}{\tau_\Pi} \frac{3\pi_s^2}{2}\, ,  \label{bulkBj}\\
D\pi_s + \frac{\pi_s}{\tau_\pi} &= \frac{4\eta}{3\tau_\pi\tau} - \left( \frac{1}{3}\tau_{\pi\pi}
+\delta_{\pi\pi}\right)\frac{\pi_s}{\tau} + \frac{2}{3}\lambda_{\pi\Pi}\frac{\Pi}{\tau} + \frac{\phi_6}{\tau_\pi} \pi_s \Pi- \frac{\phi_7}{\tau_\pi} \frac{\pi_s^2}{2} \, . \label{shearBj}
\end{align}
They have to be considered together with Eq.~(\ref{emconsBjorken}).
The transport coefficients do not depend on the anisotropy or the symmetry of the expansion. We can therefore directly compare the transport coefficients derived in~\cite{Tinti:2014cpa} with other results presented in the literature.  This is done in Figs.~\ref{fig:transport_coefficients1} and \ref{fig:transport_coefficients2}.

In the left panel of Fig.~\ref{fig:transport_coefficients1} the coefficients $\tau_{\pi\pi}, \delta_{\pi\pi}$ and $\lambda_{\pi\Pi}$ from Ref.~\cite{Jaiswal:2013npa} and \cite{Tinti:2014cpa} are plotted. They appear in the linearised equation for the shear tensor. We observe signficant differences between the results derived in \cite{Jaiswal:2013npa} and \cite{Tinti:2014cpa}, in particular, for the coefficient $\lambda_{\pi\Pi}$, which diverges in the high-temperature limit of the linearised version of anisotropic hydrodynamics. In the right panel of Fig.~\ref{fig:transport_coefficients1}  the transport coefficients $\lambda_{\Pi\pi}$ and  $\delta_{\Pi\Pi}$ from Ref.~\cite{Jaiswal:2013npa}  and from \cite{Tinti:2014cpa} are plotted. They appear correspondingly in the linearised version of the equation for the bulk pressure. Again, one observes noticeable differences between the results derived in \cite{Jaiswal:2013npa} and \cite{Tinti:2014cpa}. 

In the left (right) panel of Fig.~\ref{fig:transport_coefficients2} we present the kinetic coefficients $\phi_6$ and $\phi_7$  ($\phi_1$ and $\phi_3$) used in the linearised equation for $\pi^{\langle \mu\nu \rangle}$ ($\Pi$). All these results follow from the linearised version of anisotropic hydrodynamics \cite{Tinti:2014cpa}. In the analogous second-order hydrodynamics formulations all the $\phi$ coefficients vanish for the collision term treated in the relaxation time approximation considered in  \cite{Jaiswal:2013npa}.

Interestingly, despite of the differences between the kinetic coefficients obtained in~\cite{Tinti:2014cpa} and \cite{Jaiswal:2013npa}, the final results for the pressure anisotropy and the bulk pressure are quite similar,  see Figs.~\ref{fig:LinearPL2PT} and \ref{fig:LinearBULK}, particularly if one uses the zeroth moment in the bulk pressure evolution equation.  This behaviour may be connected with the values of the kinetic coefficients $\phi_1$ and $\phi_3$ which have noticeably smaller values for the zeroth moment case.   It is also interesting to compare Fig.~\ref{fig:NONLinearPL2PT} with Fig.~\ref{fig:LinearPL2PT}, and Fig.~\ref{fig:NONLinearBULK} with Fig.~\ref{fig:LinearBULK}. This comparison indicates that it it is  generally necessary to take into account the full equations of anisotropic hydrodynamics in order to have the best agreement with the exact solution of the kinetic equation.

\section{Summary and conclusions}
\label{sect:con}

Our results show that the use of the recently formulated leading-order anisotropic hydrodynamics is comparable with the present best prescriptions of the standard second-order viscous hydrodynamics. In specific applications, we find that the behavior of the bulk viscous pressure is better described if the zeroth moment of the kinetic equation is used, while the ratio of the longitudinal to transverse pressure is better reproduced if one uses the trace of the second moment or the projection of the second moment on the flow vector. Our results, taken together, suggest that the overall best option for the (3+1)D anisotropic hydrodynamics formulated in Ref.~\cite{Tinti:2014cpa} is to use the zeroth moment of the Boltzmann equation to derive the bulk viscous dynamics. Looking at both the close-to-equilibrium and fully non-linear results we observe that the use of the non-linear version generally improves the agreement with the exact solutions. This also suggests that the equations derived in Ref.~\cite{Tinti:2014cpa} may be successfully used as the first term in the expansion of (3+1)D anisotropic hydrodynamics.

\acknowledgments

L.T.  and W.F. were supported by the Polish National Science Center grant No. DEC-2012/06/A/ST2/00390.
R.R was supported by the Polish National Science Center grant No. DEC-2012/07/D/ST2/02125.
M.S. was supported in part by the U.S. Department of Energy, Office of Science, Office of Nuclear Physics under Awards No.~DE-SC0013470 and (within the framework of the JET Collaboration) No.~DE-AC0205CH11231. M.S. thanks the Kavli Institute for Theoretical Physics China and the Chinese Academy of Sciences under the auspices of the program ``sQGP and extreme QCD''.


\bigskip

\begin{thebibliography}{20}


\bibitem{Israel:1976tn} 
W.~Israel, Annals Phys.\  {\bf 100}, 310 (1976).

\bibitem{Israel:1979wp} 
W.~Israel and J.~M.~Stewart, Annals Phys.\  {\bf 118}, 341 (1979).
   
\bibitem{Muronga:2001zk} 
A.~Muronga,Phys.\ Rev.\ Lett.\  {\bf 88}, 062302 (2002)[Erratum-ibid.\  {\bf 89}, 159901 (2002)].
   
\bibitem{Muronga:2003ta} 
A.~Muronga, Phys.\ Rev.\ C {\bf 69}, 034903 (2004).
  
\bibitem{Baier:2006um} 
R.~Baier, P.~Romatschke and U.~A.~Wiedemann,
Phys.\ Rev.\ C {\bf 73}, 064903 (2006).
    
\bibitem{Baier:2007ix} 
R.~Baier, P.~Romatschke, D.~T.~Son,
A.~O.~Starinets and M.~A.~Stephanov,
JHEP {\bf 0804}, 100 (2008).
  
\bibitem{Romatschke:2007mq} 
P.~Romatschke and U.~Romatschke, Phys.\ Rev.\ Lett.\  {\bf 99}, 172301 (2007).
  
\bibitem{Dusling:2007gi} 
K.~Dusling and D.~Teaney,
Phys.\ Rev.\ C {\bf 77}, 034905 (2008).
  
\bibitem{Luzum:2008cw} 
M.~Luzum and P.~Romatschke,
Phys.\ Rev.\ C {\bf 78}, 034915 (2008)
[Erratum-ibid.\ C {\bf 79}, 039903 (2009)].
  
\bibitem{Song:2008hj} 
H.~Song and U.~W.~Heinz,
J.\ Phys.\ G {\bf 36}, 064033 (2009).
  
\bibitem{El:2009vj} 
A.~El, Z.~Xu and C.~Greiner,
Phys.\ Rev.\ C {\bf 81}, 041901 (2010).
  
\bibitem{PeraltaRamos:2010je} 
J.~Peralta-Ramos and E.~Calzetta,
Phys.\ Rev.\ C {\bf 82}, 054905 (2010).

\bibitem{Denicol:2010tr} 
G.~S.~Denicol, T.~Kodama and T.~Koide,
J.\ Phys.\ G {\bf 37}, 094040 (2010).
  
\bibitem{Denicol:2010xn} 
G.~S.~Denicol, T.~Koide and D.~H.~Rischke,
Phys.\ Rev.\ Lett.\  {\bf 105}, 162501 (2010).
  
\bibitem{Schenke:2010rr} 
B.~Schenke, S.~Jeon and C.~Gale,
Phys.\ Rev.\ Lett.\  {\bf 106}, 042301 (2011).
  
\bibitem{Schenke:2011tv} 
B.~Schenke, S.~Jeon and C.~Gale,
Phys.\ Lett.\ B {\bf 702}, 59 (2011).
    
\bibitem{Bozek:2009dw} 
P.~Bozek, Phys.\ Rev.\ C {\bf 81}, 034909 (2010).
   
\bibitem{Bozek:2011wa} 
P.~Bozek, Phys.\ Lett.\ B {\bf 699}, 283 (2011).
  
\bibitem{Niemi:2011ix}
H.~Niemi, G.~S.~Denicol, P.~Huovinen, E.~Molnar and D.~H.~Rischke, Phys.\ Rev.\ Lett.\  {\bf 106} (2011) 212302.
  
\bibitem{Niemi:2012ry} 
H.~Niemi, G.~S.~Denicol, P.~Huovinen, E.~Molnar and D.~H.~Rischke,    Phys.\ Rev.\ C {\bf 86}, 014909 (2012).
  
\bibitem{Bozek:2012qs} 
P.~Bozek and I.~Wyskiel-Piekarska, Phys.\ Rev.\ C {\bf 85}, 064915 (2012).
  
\bibitem{Denicol:2012cn} 
  G.~S.~Denicol, H.~Niemi, E.~Molnar and D.~H.~Rischke,
   Phys.\ Rev.\ D {\bf 85}, 114047 (2012).
  
\bibitem{Jaiswal:2013npa} 
  A.~Jaiswal,
   Phys.\ Rev.\ C {\bf 87}, 051901 (2013).
   
\bibitem{Denicol:2014loa} 
  G.~S.~Denicol,
  J.\ Phys.\ G {\bf 41}, 124004 (2014).

 
\bibitem{Florkowski:2010cf} 
  W.~Florkowski and R.~Ryblewski,
    Phys.\ Rev.\ C {\bf 83}, 034907 (2011).
  
  \bibitem{Martinez:2010sc} 
  M.~Martinez and M.~Strickland,
   Nucl.\ Phys.\ A {\bf 848}, 183 (2010).
    
\bibitem{Ryblewski:2010bs}
  R.~Ryblewski and W.~Florkowski,
   J.\ Phys.\ G {\bf 38}, 015104 (2011) .
   
\bibitem{Martinez:2010sd} 
  M.~Martinez and M.~Strickland,
    Nucl.\ Phys.\ A {\bf 856}, 68 (2011).
  
\bibitem{Ryblewski:2011aq} 
  R.~Ryblewski and W.~Florkowski,
    Eur.\ Phys.\ J.\ C {\bf 71}, 1761 (2011).
    
\bibitem{Martinez:2012tu} 
  M.~Martinez, R.~Ryblewski and M.~Strickland,
  Phys.\ Rev.\ C {\bf 85}, 064913 (2012).
  
\bibitem{Ryblewski:2012rr} 
  R.~Ryblewski and W.~Florkowski,
    Phys.\ Rev.\ C {\bf 85}, 064901 (2012).
   
\bibitem{Ryblewski:2013jsa} 
R.~Ryblewski, J.\ Phys.\ G {\bf 40}, 093101 (2013).
   
  
\bibitem{Florkowski:2012as} 
  W.~Florkowski, R.~Maj, R.~Ryblewski and M.~Strickland, Phys.\  Rev.\  C 87, {\bf 034914} (2013).
  
\bibitem{Florkowski:2013uqa} 
  W.~Florkowski and R.~Maj,
  Acta Phys.\ Polon.\ B {\bf 44}, no. 10, 2003 (2013).
  
\bibitem{Florkowski:2014txa} 
  W.~Florkowski and O.~Madetko,
  Acta Phys.\ Polon.\ B {\bf 45}, 1103 (2014)
  
  
\bibitem{Romatschke:2003ms} 
  P.~Romatschke and M.~Strickland,
  Phys.\ Rev.\ D {\bf 68}, 036004 (2003). 
  
  
\bibitem{Bazow:2013ifa}
  D.~Bazow, U.~W.~Heinz and M.~Strickland,
  Phys.\ Rev.\ C {\bf 90}, no. 5, 054910 (2014).
   
\bibitem{Bazow:2015cha} 
  D.~Bazow, U.~W.~Heinz and M.~Martinez,
  arXiv:1503.07443 [nucl-th].
   

\bibitem{Tinti:2013vba} 
  L.~Tinti and W.~Florkowski,
  Phys.\ Rev.\ C {\bf 89}, no. 3, 034907 (2014).
  
\bibitem{Nopoush:2014pfa} 
  M.~Nopoush, R.~Ryblewski and M.~Strickland,
  Phys.\ Rev.\ C {\bf 90}, no. 1, 014908 (2014).

\bibitem{Tinti:2014cpa} 
  L.~Tinti,
  arXiv:1411.7615 [hep-ph].
  

\bibitem{Florkowski:2013lza} 
  W.~Florkowski, R.~Ryblewski and M.~Strickland,
  Nucl.\ Phys.\ A {\bf 916}, 249 (2013).
  
\bibitem{Florkowski:2013lya} 
  W.~Florkowski, R.~Ryblewski and M.~Strickland,
  Phys.\ Rev.\ C {\bf 88}, 024903 (2013).
  
  \bibitem{Denicol:2014xca} 
  G.~S.~Denicol, U.~W.~Heinz, M.~Martinez, J.~Noronha and M.~Strickland,
  Phys.\ Rev.\ Lett.\  {\bf 113}, 202301 (2014).
  
  \bibitem{Denicol:2014tha} 
  G.~S.~Denicol, U.~W.~Heinz, M.~Martinez, J.~Noronha and M.~Strickland,
  Phys.\ Rev.\ D {\bf 90}, 125026 (2014).

\bibitem{Nopoush:2014qba} 
  M.~Nopoush, R.~Ryblewski and M.~Strickland,
  Phys.\ Rev.\ D {\bf 91}, no. 4, 045007 (2015).
  
     
  \bibitem{Denicol:2014mca} 
  G.~S.~Denicol, W.~Florkowski, R.~Ryblewski and M.~Strickland,
  Phys.\ Rev.\ C {\bf 90}, no. 4, 044905 (2014).
  
  \bibitem{Jaiswal:2014isa} 
  A.~Jaiswal, R.~Ryblewski and M.~Strickland,
  Phys.\ Rev.\ C {\bf 90}, no. 4, 044908 (2014).
  
    
\bibitem{Florkowski:2014sfa} 
  W.~Florkowski, E.~Maksymiuk, R.~Ryblewski and M.~Strickland,
  Phys.\ Rev.\ C {\bf 89}, no. 5, 054908 (2014).
  
\bibitem{Florkowski:2014bba} 
  W.~Florkowski, R.~Ryblewski, M.~Strickland and L.~Tinti,
  Phys.\ Rev.\ C {\bf 89}, no. 5, 054909 (2014).
  

  
  
   

\bibitem{Florkowski:2014sda} 
  W.~Florkowski and E.~Maksymiuk,
  J.\ Phys.\ G {\bf 42}, no. 4, 045106 (2015).

\bibitem{Florkowski:2015lra} 
  W.~Florkowski, A.~Jaiswal, E.~Maksymiuk, R.~Ryblewski and M.~Strickland,
  arXiv:1503.03226 [nucl-th].

  

  
  
\bibitem{Bhatnagar:1954zz} 
  P.~L.~Bhatnagar, E.~P.~Gross and M.~Krook,
   Phys.\ Rev.\  {\bf 94}, 511 (1954).
  

\bibitem{Florkowski:2011jg} 
  W.~Florkowski and R.~Ryblewski,
    Phys.\ Rev.\ C {\bf 85}, 044902 (2012).


\bibitem{Anderson:1974} 
  J.~L.~Anderson and  H.~R.~Witting,
   Physica  {\bf 74}, 466 (1974).

\bibitem{Czyz:1986mr} 
  W.~Czyz and W.~Florkowski,
   Acta Phys.\ Polon.\ B {\bf 17}, 819 (1986).



\end{thebibliography}
\end{document}